\documentclass{article}
\usepackage{graphicx}
\begin{document}
\title{ The $O(n)$ Model in the $n\rightarrow 0$ Limit ( self-avoiding-walks)\\ and \\ Logarithmic Conformal
Field Theory}
\author{M. Sadegh Movahed$^1$\footnote{e-mail: m.s.movahed@mehr.sharif.edu} , M. Saadat$^1$\footnote{e-mail: msaadat@sharif.edu} and M. Reza Rahimi Tabar$^{1,2}$ \footnote{e-mail: rahimitabar@iust.ac.ir
}\\ \\
{\small  \it $^1$Department of Physics, Sharif University of
Technology,
 Tehran, P.O.Box: 11365-9161, Iran}\\
{\small  \it $^2$CNRS UMR 6529, Observatoire de la C$\hat o$te
d'Azur, BP 4229, 06304 Nice Cedex 4, France }}
\date{}
\maketitle

\def\p{\partial}
\def\be{\begin{equation}}
\def\bea{\begin{eqnarray}}
\def\ee{\end{equation}}
\def\eea{\end{eqnarray}}
\def\bearst{\begin{eqnarray*}}
\def\eearst{\end{eqnarray*}}
\def\dbar{\bar \partial}
\def\nn{\nonumber}
\def\ll{\lambda}
\def\l{\label}
\def\D{\Delta}
\def\o{\over}
\def\E{{\rm e}}
\def\peleven{\parbox{14cm}}
\def\peffec{\peight{\bearst\eearst}\hfill\peleven}
\def\pspace{\peight{\bearst\eearst}\hfill}
\def\ptwelve{\parbox{15cm}}
\def\peight{\parbox{8mm}}
        \def\sl{\hbox{{\rm sl}}}
     \def\half{{1 \over 2}}
     \def\Uq#1{{\rm U}_q \left( #1 \right) }
     \def\Um#1{{\rm U}_{\mu} \left( #1 \right) }
     \def\U#1{{\rm U} \left( #1 \right) }
     \def\D{{\Delta }}
     \def\P{{\Phi}}
     \def\p{{\phi}}
     \def\si{{\psi}}
     \def\d{{\partial}}
     \def\R{{\bf R}}
     \def\s{\sum}
     \def\pr{\prod}
     \def\a{\alpha}
     \def\eps{\epsilon}
     \def\HG{\H_{"G"}}
     \def\Ts{\tilde{\psi}}
     \def\Si{\Psi}
     \def\Raw{\Rightarrow}
     \def\raw{\rightarrow}
     \def\da{\dag}
     \def\lm{\lambda}
     \def\sq{\sqrt}
     \def\D{\Delta}
     \def\b{\beta}
     \def\i{\imath}
     \def\im{\rm {\it Im}}
     \def\re{\rm {\it Re}}
     \def\ah{\hat{A}}
     \def\bh{\hat{B}}
     \def\ch{\hat{C}}
     \def\eh{\hat{D}}
     \def\c{\cdot}
     \def\cs{\cdots}
     \def\e{\epsilon}
     \def\g{\gamma}
\def\CN{{\cal N}}
\def\CM{{\cal M}}
\def\CA{{\cal A}}
\def\CB{{\cal B}}
\def\CD{{\cal D}}
\def\CO{{\cal O}}
\def\CD{{\cal D}}
\def\W{{\cal W}}
\def\lk{\left [}
\def\rk{\right ]}
\def\kt#1{\mid{{#1}}>}
\def\br#1{<{#1}\mid}
\def\z#1{z_{#1}}
\def\de#1{\Delta_{#1}}
\def\ds{\Delta_\Psi}
\def\df{\Delta_\Phi}
\def\da{\Delta_A}
\def\db{\Delta_B}
\def\dc{\Delta_p}
\def\I{\rm {I\kern-.3em I}}
\def\C{\rm {I\kern-.520em C}}
\def\R{\rm {I\kern-.3em R}}
\def\CZ{\rm {Z\kern-.4em Z}}
\def\str{$sl(2,\C)$}
 \def\ps#1{\psi({#1})}
\def\unit{\rm {1\kern-.4em 1}}
\def\f{\frac}
    \def\ff{\f{1}{2}}
    \def\pp{<\Psi(\z1)\Psi(\z2)>}
\def\abf{<A(z_1)B(z_2)\Phi(z_3)>}
\def\abs{<A(z_1)B(z_2)\Psi(z_3)>}
\def\ab#1{\vert #1 \vert}
\def\kp{\kt{\de{p}}}
\def\kpp{\kt{{\D'}_p}}
\def\kk{{\bf k}}
\def\E{{\rm e}}
\def\peleven{\parbox{11cm}}
\def\peffec{\peight{\bearst\eearst}\hfill\peleven}
\def\pspace{\peight{\bearst\eearst}\hfill}
\def\ptwelve{\parbox{12cm}}
\def\peight{\parbox{8mm}}

\begin{abstract}
We consider the $O(n)$ theory in the $n \rightarrow 0$ limit. We
show that the theory is described by logarithmic conformal field
theory, and that the correlation functions have logarithmic
singularities. The explicit forms of the two-, three- and
four-point correlation functions of the scaling fields and the
corresponding logarithmic partners are derived.
\end{abstract}

\section{Introduction}
The $n$-vector model with its Landau-Ginzburg-Willson Hamiltonian,
which has $O(n)$ symmetry, may be used to study physical
properties of many critical systems. For example, in the limit
$n=1$ we obtain Ising-like systems which describe liquid-vapor
transitions in the classical and critical binary fluids. The
Helium superfluid transition corresponds to the limit $n=2$. Only
for the case of $n=3$ does the experimental information come from
truly ferromagnetic systems. The limit $n \to 0$ describes the
statistical properties of self-avoiding walks (SAWs), which
describe the universal properties of linear polymers, i.e., long
nonintersecting chains, in a dilute solution. These properties can
be computed by such techniques as the transfer-matrix method
\cite{guim}, series expansions \cite{zhao}, and the scanning Monte
Carlo method \cite{mei}. More generally, deep insight into the
problem of computing the statistical properties of linear polymers
and the SAWs can be obtained from the $O(n)$ model in the $n \to
0$ limit, a discovery first made by de Gennes
\cite{gennes1,gennes2}.

Recently, it has been shown by Cardy that the $O(n)$ model in the
$n\rightarrow 0$ limit has a logarithmic conformal field
structure, with its correlation function having logarithmic
singularity [7]. The logarithmic conformal field theories (LCFT)
[6,7] are extensions of the conventional conformal field theories
(CFT) [8-12], which have emerged in recent years in a number of
interesting physical problems, such as the WZNW models [13-18],
supergroups and super-symmetric field theories [19-25],
Haldane-Rezzayi state in the fractional quantum Hall effect
[26-30], multi-fractality [31], two-dimensional turbulence
[32-34], gravitaitionally-dressed theories [35], polymer and
abelian sandpiles [36-40], string theory [41-44] and the D-brane
recoil [45-55], Ads/CFT correspondence [56-67], Seiberg-Witten
solution to SUSY Yang-Mills theory [68] and disordered systems
[69-80]. Moreover, such related issues as the Null vectors,
Characters, partition functions, fusion rules, Modular Invariance,
C-theorem, LCFT`s with boundary and operator product expansions
have been discussed in Refs. [81-120].

The LCFT are characterized by the fact that their dilatation
operators $L_0$ are not diagonalized and admit a Jordan cell
structure. The non-trivial mixing between these operators leads to
logarithmic singularities in their correlation functions. It has
been shown in Ref. [8] that the correlator of two fields in such
field theories has a logarithmic singularity as follows:

 \be
\langle\psi(r_1) \psi(r_2)\rangle \sim {|r_1-r_2|}^{-2
\Delta_{\psi}} \ln|r_1-r_2| + \ldots \ee

In this paper, we consider the correlation functions of the
scaling fields of the $O(n)$ model and derive explicit expressions
for the two-, three- and four-point correlation functions. In
Section 2, we derive the scaling fields and their scaling
exponents. In Section 3, we derive the Jordan cell structure of
the theory in the $n\rightarrow 0$ limit, while Section 4 presents
the derivation of the two-, three- and four-point correlation
functions of the scaling fields and their logarithmic partners.
The details of calculations are presented in the Appendices.
\section{Scaling fields in the $O(n)$ model}
The Landau-Ginzburg description of the $O(n)$ model starts with
the effective Hamiltonian: \be\label{hamiltonian}
H=\int\frac{1}{2}\sum_{a}:[(\nabla\phi_{a})^{2}+m^{2}\phi_{a}^{2}]:+g\sum_{a,b}:\phi_{a}^{2}\phi_{b}^{2}:
d^{d}r, \ee where $a,b=1,2,\cdots,n$, and $g$ is the coupling
constant of the perturbation,
$\sum_{a,b}:\phi_{a}^{2}\phi_{b}^{2}:$, to the free model. Suppose
that under a scaling transformation
$\vec{r}\rightarrow\vec{\acute{r}}=\lambda\vec{r}$, $\phi_a$
behaves as,
$\phi_a\rightarrow\acute{\phi_a}=\lambda^{-x_{\phi_a}^{0}}\phi_a$,
where $x_{\phi_a}^{0}$ is called the scaling dimension of the
field $\phi_a$. Invariance of the Hamiltonian under scaling
requires the coupling $g$ to have the scaling dimension,
$y_{g}=d-4x_{\phi_a}^{0}$. It would be relevant at the pure fixed
point if, $d>4x_{\phi_a}^{0}$ [120-121]. If $y_g$ is small, it is
possible to develop a perturbative renormalization group (RG)
equation in powers of these variables, which can then yield the
fixed points.

To develop the RG equation for a typical coupling constant
$g_{i}$, we need the operator product expansion (OPE)
coefficients. The general form of the $beta$-function for coupling
$g_i$ is given by [6]: \bea\label{beta}
\beta_{g_{i}}\equiv\frac{dg_{i}}{dl}=y_{g_{i}}g_{i}-\sum_{jk}C_{ijk}g_{j}g_{k}+\cdots,
\eea where $l>1$ is a re-scaling parameter. To derive the OPE
coefficients we note that, when $m^2=g=0$, we obtain the Gaussian
model, i.e.,
$H=\frac{1}{2}\int\sum_{a}:(\nabla\phi_{a})^{2}:d^{d}r$. In the
Gaussian model, the various components $\phi_{a}$ are decoupled,
so that the two-point correlation function has the following form:
\begin{eqnarray}
\langle\phi_{a}(r_{i})\phi_{b}(r_{j})\rangle=\frac{\delta_{ab}}{r_{ij}^{d-2}},
\end{eqnarray}
where, $r_{ij}=|\vec{r}_{i}-\vec{r}_{j}|$. Considering
$\Phi=\sum_{a,b}:\phi_{a}^{2}\phi_{b}^{2}:$ as the perturbative
term in $O(n)$ Hamiltonian, and using Wick's theorem, one can
evaluate the OPE of the field $\Phi$ with itself as: \bea
\Phi\cdot\Phi&=&(\sum_{a,b}:\phi_a^2\phi_b^2:)\cdot(\sum_{c,d}:\phi_c^2\phi_d^2:),\nonumber\\
&=&24n^2+96nE+\xi_{\Phi}\Phi+\cdots, \eea where,
$E=\sum_{c}:\phi_{c}^{2}:$, $n=\sum_{ab}\delta_{ab}$ and
$\xi_{\Phi}=(8n+64)$ (see Appendix A). Therefore, using Eq.(3),
one obtains the following RG equation for $g$: \bea \label{6}
\beta_{g}=y_{g}g-\xi_{\Phi}g^{2}+\cdots. \eea To check that the
field $E$ is a scaling operator, we consider the OPE of $E$ with a
scaling dimension $x_{E}(n)$ and $\Phi$: \bea
E\cdot\Phi&=&(\sum_{c}:\phi_{c}^{2}:).(\sum_{a,b}:\phi_a^2\phi_b^2:),\nonumber\\
&=&8\Phi+\xi_{E}E+\cdots, \eea where, $\xi_{E}=(4n+8)$. Therefore,
this form of the OPE for $E\cdot\Phi$ shows that the field $E$ is
a scaling operator.

As noted by Cardy [7], there is also another scaling field in the
theory, $\widetilde{E}_{ab}$, with a scaling dimension
$x_{\widetilde{E}}(n)$ which is given explicitly by the following
expression in terms of the fields $\phi_a$:
$\widetilde{E}_{ab}=:\phi_{a}\phi_{b}:-\frac{\delta_{ab}}{n}\sum_{c}:\phi_{c}^{2}:$.
Its OPE with $\Phi$ is given by the following expression: \bea
\widetilde{E}_{ab}.\Phi &=& (:\phi_{a}\phi_{b}:-\frac{\delta_{ab}}{n}\sum_{c}:
\phi_{c}^{2}:).(\sum_{d,e}:\phi_d^2\phi_e^2:),\nonumber\\
&=&-\frac{8}{n}\delta_{ab}\Phi+\xi_{\widetilde{E}}\widetilde{E}_{ab}+\cdots,
\eea with $\xi_{\widetilde{E}}=8$.

To derive the scaling dimensions of the fields $E$ and
$\widetilde{E}_{ab}$, we perturb the Gaussian Hamiltonian by these
fields with  the coupling $t_1$ and $t_2$, respectively. Then, it
is clear that their scaling dimensions are: \bea\label{betat1}
y_{t_{1}}^{0}&=&d-x_{E}^{0},\nonumber\\
y_{t_{2}}^{0}&=&d-x_{\widetilde{E}}^{0}. \eea Then, according to
Eq.(\ref{beta}), the RG equations for the couplings are given by:
\bea
\beta_{t_{1}}&=&y_{t_{1}}t_{1}-2\xi_{E}gt_{1}+\cdots,\nonumber\\
\beta_{t_{2}}&=&y_{t_{2}}t_{2}-2\xi_{\widetilde{E}}gt_{2}+\cdots
\eea Because the fixed points and the RG eigenvalues correspond to
the "zeros" and derivatives of the RG $beta$ function at the fixed
points, respectively, one has: \bea
\label{9}g^{*}=\frac{y^{0}_{g}}{\xi_{\Phi}}. \eea Using
Eqs.(\ref{betat1}), we obtain: \bea \label{4}
y_{t_{1}}&=&\left.\frac{\partial\beta_{t_{1}}}{\partial
t_{1}}\right|_{0},\nonumber\\
&=&y_{t_{1}}^{0}-2y_{g}^{0}\frac{\xi_{E}}{\xi_{\Phi}}, \eea and,
similarly: \bea
y_{t_{2}}&=&y_{t_{2}}^{0}-2y_{g}^{0}\frac{\xi_{\widetilde{E}}}{\xi_{\Phi}}.
\eea Substituting the values of $\xi$'s, $y_{t_{1}}^{0}$, and
$y_{t_{2}}^{0}$, and denoting $y_{t_{1}}=d-x_{E}$ and
$y_{t_{2}}=d-x_{\widetilde{E}}$, we obtain the following scaling
dimensions for $E$ and $\widetilde{E}$: \bea
x_{E}(n)&=&x_{E}^{0}-\frac{2(4n+8)}{(8n+64)}y_{g}^{0}+\cdots,\nonumber\\
x_{\widetilde{E}}(n)&=&x_{\widetilde{E}}^{0}-\frac{16}{(8n+64)}y_{g}^{0}+\cdots.
\eea One can then derive the OPE of the field $:\phi_{a}\phi_{b}:$
with the scaling dimension $x_{\phi\phi}(n)$ and $\Phi$. If this
is done, one obtains: \be
:\phi_{a}\phi_{b}:.\Phi=4\delta_{ab}E+\xi_{\widetilde{E}}:\phi_{a}\phi_{b}:+\cdots,
\ee Therefore, it is clear that $:\phi_{a}\phi_{b}:$ and
$\widetilde{E}$ have the same scaling dimension.
\section{Scaling dimensions in the $n \to 0$  Limit}
It is evident that in the $n\to 0$ limit, Eqs.(14) reduce to the
following: \bea
x_{E}(0)&=&x_{E}^{0}-\frac{1}{4}y_{g}^{0}+O(y_{g}^{0})^2,\nonumber\\
x_{\phi\phi}(0)=
x_{\widetilde{E}}(0)&=&x_{\widetilde{E}}^{0}-\frac{1}{4}y_{g}^{0}+O(y_{g}^{0})^2.
\eea Therefore, in the $n\rightarrow0$ limit, there are three
fields with the same scaling dimensions. It is well-known in CFT
that two scaling fields with the same conformal weights may
constitute a Jordan cell of rank 2. Such theories are what are
referred to as the logarithmic conformal field theories (LCFT).
The correlation functions in LCFT may have logarithmic as well as
power-law terms. Before calculating these correlation functions,
we can show that $E(r)$ and $:\phi_{a}\phi_{b}(r):$ constitute a
Jordan cell in the $n \to 0$ limits. Since $E(r)$ and
$\widetilde{E}(r)$ are two scaling fields in the $O(n)$ model with
scaling dimensions $x_{E}(n)$ and $x_{\widetilde{E}}(n)$, we have:
\begin{eqnarray}\label{scale}
E(\lambda r)&=&\lambda^{-x_{E}(n)}E(r),\nonumber\\
\widetilde{E}(\lambda
r)&=&\lambda^{-x_{\widetilde{E}}(n)}\widetilde{E}(r).
\end{eqnarray}
We are interested in the scaling behavior of
$:\phi_{a}\phi_{b}(r):$ in the $n\to 0$ limit. First, we write
$:\phi_{a}\phi_{b}(r):$ in terms of $E(r)$ and
$\widetilde{E}_{ab}(r)$ as:
\begin{eqnarray}\label{fifi}
:\phi_{a}\phi_{b}(r):=\widetilde{E}_{ab}(r)+\frac{\delta_{ab}}{n}E(r),
\end{eqnarray}
and, then, using Eqs.(\ref{scale}), we find that:
\begin{eqnarray}
:\phi_{a}\phi_{b}(\lambda
r):=\lambda^{-x_{E}(n)}[\lambda^{x_{E}(n)-x_{\widetilde{E}}(n)}\widetilde{E}_{ab}(r)+\frac{\delta_{ab}}{n}E(r)].
\end{eqnarray}
Since, $x_{E}(0)=x_{\widetilde{E}}(0)$, then:
\begin{eqnarray}\label{taylor}
\lambda^{x_{E}(n)-x_{\widetilde{E}}(n)}=1+nu\ln\lambda+O(n^{2})+\cdots,
\end{eqnarray}
where, $u=x^{\prime}_{E}(0)-x^{\prime}_{\widetilde{E}}(0)$. Now,
it is straightforward to check that:
\begin{eqnarray}
\lim_{n\rightarrow0}:\phi_{a}\phi_{b}(\lambda
r):=\lambda^{x_{E}(0)}[:\phi_{a}\phi_{b}(r):-\ln\lambda(u\delta_{ab}E(r))].
\end{eqnarray}
Therefore, $u\delta_{ab}E(r)$ and $:\phi_{a}\phi_{b}(r):$ are
degenerate fields in the $n\rightarrow0$ limit which form a Jordan
cell:
\begin{eqnarray}
\left(\begin{array}{c}
u\delta_{ab}E(r) \\
\\
:\phi_{a}\phi_{b}(r):
\end{array}
\right).
\end{eqnarray}
\section{Two, Three and four point correlation functions}
In this section we first derive the two-point correlation
functions of the fields $\widetilde{E}_{ab}(r)$ and $E(r)$, for
arbitrary $n$. As shown in Section 2, these fields have the
scaling dimensions $x_{\widetilde{E}}(n)$ and $x_{E}(n)$,
respectively. However, to fix the amplitude of the two-point
correlation function and their tensorial structure we note that:
 \bea\langle E(r_1)E(r_2)\rangle=\langle
 (\sum_{i}:\phi_{i}^{2}(r_1):)(\sum_{j}:\phi_{j}^{2}(r_2):)\rangle
 ,\eea
where normal ordered is defined as  $ :: = 1 - <  > $.
 The two-point correlation function of the field $\widetilde{E}_{ab}(r)$ is given by:
 \be\langle \widetilde{E}_{ab}(r_1)\widetilde{E}_{cd}(r_2)\rangle = \langle(:\phi_{a}\phi_{b}(r_1):-\frac{\delta_{ab}}{n}\sum_{i}:\phi_{i}^{2}(r_1):)
(:\phi_{c}\phi_{d}(r_2):-\frac{\delta_{cd}}{n}\sum_{j}:\phi_{j}^{2}(r_2):)\rangle
, \ee and similarly for $\langle
\widetilde{E}_{ab}(r_1)E(r_2)\rangle$.
The right-hand side of the above equations can be evaluated using
Wick's theorem and noting that
$\langle\phi_{a}\phi_{b}\rangle\sim\delta_{ab}$. Therefore:
\bea\langle:\phi_{a}\phi_{b}::\phi_{c}\phi_{d}:\rangle
&\sim&D_{ab,cd},\nonumber\\
\sum_{i}\langle:\phi_{a}\phi_{b}::\phi_{i}^{2}:\rangle&\sim&2\delta_{ab},\nonumber\\
\sum_{ij}\langle:\phi_{i}^2::\phi_{j}^{2}:\rangle&\sim&2n, \eea
where $D_{ab,cd}=\delta_{ac}\delta_{bd}+\delta_{ad}\delta_{bc}$.
However, we have fixed their tensorial structures by calculation
of the two-point correlation function using the free Hamiltonian.
Indeed, the interaction will change the amplitude and the scaling
exponent, but dose not affect
the tensorial structures.\\
 Using the Eqs.(25), we find that: \bea
 \langle \widetilde{E}_{ab}(r_1)E(r_2)\rangle&=&0,\nonumber\\
\langle E(r_1)E(r_2)\rangle&=&2nA(n)r_{12}^{-2x_{E}(n)},\nonumber\\
 \langle
\widetilde{E}_{ab}(r_1)\widetilde{E}_{cd}(r_2)\rangle&=&[D_{ab,cd}-\frac{2}{n}\delta_{cd}\delta_{ab}]\widetilde{A}(n)r_{12}^{-2x_{\widetilde{E}}(n)},\eea
where $A(n)$ and $\widetilde{A}(n)$ are two functions have Taylor
expansion at $n=0$.

 We can now derive two-point correlation functions of fields
$E$ (scaling operator) and $:\phi_{a}\phi_{b}:$ (logarithmic
operator) in the $n \to 0$ limit. Singularity of the two point
functions at $n\rightarrow0$ limit can be removed by choosing
$A(0)=\widetilde{A}(0)$, as has recently been shown  by Cardy [7]:
\bea \lim_{n \rightarrow0}\langle
E(r_{1})E(r_{2})\rangle&=&0,\nonumber\\
\lim_{n \rightarrow0}\langle
:\phi_{a}\phi_{b}(r_{1}):E(r_{2})\rangle&=&2A(0)\delta_{ab}r_{12}^{-2x_{E}(0)},\\
 \lim_{n
\rightarrow0}\langle:\phi_{a}\phi_{b}(r_{1})::\phi_{c}\phi_{d}(r_{2}):\rangle&=&\left\{A(0)\lk
D_{ab,cd}-4u\delta_{ab}\delta_{cd}\ln r_{12}\rk
+2\delta_{ab}\delta_{cd}
[A^{\prime}(0)-\widetilde{A}^{\prime}(0)]\right\}r_{12}^{-2x_{E}(0)}.\nonumber
\eea  This is a general property of correlation functions of the
Jordan cell components in a LCFT [9].

  In what follows we derive various three- and four-point
correlation functions of the Jordan cell components. In the CFT,
due to the conformal symmetry, the three-point correlation
function of the scaling fields has the following form:
\begin{eqnarray}
\langle\varphi_{1}(r_{1})\varphi_{2}(r_{2})\varphi_{3}(r_{3})\rangle=\frac{C_{123}}{r_{12}^{\Delta_{1}+\Delta_{2}-\Delta_{3}}
r_{13}^{\Delta_{1}+\Delta_{3}-\Delta_{2}}r_{23}^{\Delta_{2}+\Delta_{3}-\Delta_{1}}},
\end{eqnarray}
where $\Delta_{1}$, $\Delta_{2}$ and $\Delta_{3}$ are scaling
dimensions of $\varphi_{1}$, $\varphi_{2}$ and $\varphi_{3}$,
respectively, and $C_{123}$ is a parameter that depends on the
model.

 We are interested in various three-point correlation functions of
 scaling field $E$ and $:\phi_{a}\phi_b:$ which is its logarithmic
 partner at $n \to 0$ limit. Using Eq.(\ref{fifi}) and substitute of
$:\phi_{i}\phi_{j}:$ in terms of $E$ and $\widetilde{E}$,
three-point correlation function of logarithmic field
$\langle:\phi_{a}\phi_{b}(r_{1})::\phi_{c}\phi_{d}(r_{2})::\phi_{e}\phi_{f}(r_{3}):\rangle$
can be written as follows:

\begin{eqnarray}
&\hskip -50mm\langle:\phi_{a}\phi_{b}(r_{1})::\phi_{c}\phi_{d}(r_{2})::\phi_{e}\phi_{f}(r_{3}):\rangle=A(n)(r_{12}r_{13}r_{23})^{-x_{\widetilde{E}}(n)}\nonumber\\
&\hskip 15mm
\times\left[D_{ab,cd,ef}-\frac{4}{n}(\delta_{ab}D_{cd,ef}+\delta_{cd}D_{ab,ef}+\delta_{ef}D_{ab,cd})+\frac{16}{n^2}\delta_{ab}
\delta_{cd}\delta_{ef}\right]\nonumber\\
&+\frac{\delta_{ab}}{n}\left(4D_{cd,ef}-\frac{8}{n}\delta_{cd}\delta_{ef}\right)B(n)
(r_{12}r_{13})^{-x_{E}(n)}r_{23}^{x_{E}(n)-2x_{\widetilde{E}}(n)}\nonumber\\
&+\frac{\delta_{cd}}{n}\left(4D_{ab,ef}-\frac{8}{n}\delta_{ab}\delta_{ef}\right)B(n)
(r_{12}r_{23})^{-x_{E}(n)}r_{13}^{x_{E}(n)-2x_{\widetilde{E}}(n)}\nonumber\\
&+\frac{\delta_{ef}}{n}\left(4D_{ab,cd}-\frac{8}{n}\delta_{ab}\delta_{cd}\right)B(n)
(r_{13}r_{23})^{-x_{E}(n)}r_{12}^{x_{E}(n)-2x_{\widetilde{E}}(n)}\nonumber\\
&\hskip
-31mm+\frac{1}{n^3}8n\delta_{ab}\delta_{cd}\delta_{ef}C(n)(r_{12}r_{13}r_{23})^{-x_{E}(n)},
\end{eqnarray}
where $A(n)$, $B(n)$, $C(n)$ are functions that have Taylor
expansion near $n=0$ (see appendix B for details). To obtain the
$n\rightarrow0$ limit, we expand above expression about $n=0$ by
writing down the Taylor expansion of $A(n)$, $B(n)$ and $C(n)$,
and $r_{ij}^{x_{E}(n)-x_{\widetilde{E}}(n)}$. It is not difficult
to see that resulting relation will be divergent in the
$n\rightarrow0$ limit for arbitrary values of $A(0)$, $B(0)$,
$C(0)$, $A^{\prime}(0)$, $B^{\prime}(0)$ and $C^{\prime}(0)$.
However, the divergent terms will cancel each other if we choose a
special case in which $A(0)=B(0)=C(0)$, and
$A^{\prime}(0)=B^{\prime}(0)=C^{\prime}(0)$, so that:
\begin{eqnarray}\label{sadegh}
&\hskip -50mm \lim_{n\rightarrow0}\langle:\phi_{a}\phi_{b}(r_{1})::\phi_{c}\phi_{d}(r_{2})::\phi_{e}\phi_{f}(r_{3}):\rangle =(r_{12}r_{13}r_{23})^{-x_{E}(0)}\nonumber\\
&\hskip 15mm \times\left\{A(0)\left[D_{ab,cd,ef}
-4u\left(\delta_{ab}D_{cd,ef}\ln\frac{r_{12}r_{13}}{r_{23}}+\delta_{cd}D_{ab,ef}\ln\frac{r_{12}r_{23}}{r_{13}}
+\delta_{ef}D_{ab,cd}\ln\frac{r_{13}r_{23}}{r_{12}}\right)\right.\right.\nonumber\\
&+\left.\left.8u^2\delta_{ab}\delta_{cd}\delta_{ef}[4(\ln
r_{12}\ln r_{13}+\ln r_{12}\ln r_{23}+\ln r_{13}\ln
r_{23})-\ln^{2}(r_{12}r_{13}r_{23})]\right]\right.\nonumber\\
&\hskip
-50mm+\left.\delta_{ab}\delta_{cd}\delta_{ef}\lk8A^{\prime\prime}(0)-12B^{\prime\prime}(0)+4C^{\prime\prime}(0)\rk\right\}.
\end{eqnarray}

 In the same manner other three-point correlation functions obtain
 at $n \to 0$ limit as follows:
 \bea \lim_{n \rightarrow0}\langle
E(r_{1})E(r_{2})E(r_{3})\rangle&=&0,\nonumber\\
 \lim_{n \rightarrow0}\langle :\phi_{a}\phi_{b}(r_{1}):E(r_{2})E(r_{3})\rangle&=&8A(0)\delta_{ab}(r_{12}r_{13}r_{23})^{-x_{E}(0)},\\
\lim_{n\rightarrow0}\langle:\phi_{a}\phi_{b}(r_{1})::\phi_{c}\phi_{d}(r_{2}):E(r_{3})\rangle&=&
A(0)\lk4D_{ab,cd}-16u\delta_{ab}\delta_{cd}\ln r_{12}\rk
(r_{12}r_{13}r_{23})^{-x_{E}(0)}.\nonumber\eea

  Finally, the four-point correlation functions of $E$ and $:\phi_a \phi_b:$ can be calculated just as the same as three-point correlation
 functions. Some key-functions which are useful in this calculation given in appendix B.

  In the $n \to 0$ limit we encounter divergent terms in four-point correlation functions, but the divergent terms will cancel each other if we choose
  a special case in which
  $A(0)=B(0)=C(0)=D(0)$
, $A^{\prime}(0)=B^{\prime}(0)=C^{\prime}(0)=D^{\prime}(0)$,
$A^{\prime\prime}(0)=B^{\prime\prime}(0)=C^{\prime\prime}(0)=D^{\prime\prime}(0)$
and $f_1(\eta)=f_2(\eta)=f_3(\eta)=f_4(\eta)$, Therefore:
\begin{eqnarray}
 \lim_{n\rightarrow0}\langle
E(r_{1})E(r_{2})E(r_{3})E(r_{4})\rangle&=&0,\nonumber\\
\lim_{n\rightarrow0}\langle
:\phi_{a}\phi_{b}(r_{1}):E(r_{2})E(r_{3})E(r_{4})\rangle&=&48A(0)f(\eta)\delta_{ab}(r_{12}r_{13}r_{14}r_{23}r_{24}r_{34})^{-2x_{E}(0)/3},\nonumber\\
\lim_{n\rightarrow0}\langle:\phi_{a}\phi_{b}(r_{1})::\phi_{c}\phi_{d}(r_{2}):E(r_{3})E(r_{4})\rangle&=&A(0)f(\eta)
(24D_{ab,cd}+8-16u\ln \frac{
r_{13}r_{23}r_{14}r_{24}r_{12}^{4}}{r_{34}^{2}}\delta_{ab}\delta_{cd})\nonumber\\
&&\quad\times(r_{12}r_{13}r_{14}r_{23}r_{24}r_{34})^{-2x_{E}(0)/3},
\end{eqnarray}
\begin{eqnarray}
&\hskip -50mm\lim_{n\rightarrow0}\langle:\phi_{a}\phi_{b}(r_{1})::\phi_{c}\phi_{d}(r_{2})::\phi_{e}\phi_{f}(r_{3}):E(r_{4})\rangle=\nonumber\\
&A(0)f(\eta)\left\{\frac{}{}[6D_{ab,cd,ef}+2\delta_{ab}D_{cd,ef}+2\delta_{cd}D_{ab,ef}+2\delta_{ef}D_{ab,cd}]\right.\nonumber\\
&\hskip 25mm
\left.+8u[\delta_{ab}D_{cd,ef}\ln\frac{r_{23}r_{24}r_{34}}{(r_{12}r_{13}r_{14})^{2}}+\delta_{cd}D_{ab,ef}\ln\frac{r_{13}r_{14}r_{34}}{(r_{12}r_{23}r_{24})^{2}}
+\delta_{ef}D_{ab,cd}\ln\frac{r_{12}r_{24}r_{14}}{(r_{23}r_{13}r_{34})^{2}}\right.\nonumber\\&
\hskip 15mm \left.-\delta_{ab}\delta_{cd}\delta_{ef}\ln
r_{12}r_{13}r_{23}]-16u^{2}\delta_{ab}\delta_{cd}\delta_{ef}[-2\ln
r_{12}\ln r_{34}-3\ln r_{12}\ln
r_{23}\right.\nonumber\\&\left.-2\ln r_{13}\ln r_{24}-3\ln
r_{12}\ln r_{13}-2\ln r_{23}\ln r_{14}-3\ln r_{13}\ln
r_{23}\right.\nonumber\\&\left.+\ln r_{12}\ln r_{14}r_{24}-\ln
r_{14}\ln r_{24}+\ln r_{13}\ln r_{14}r_{34}-\ln r_{14}\ln
r_{34}\right.\nonumber\\&\left.+\ln r_{23}\ln r_{24}r_{34}-\ln
r_{24}\ln
r_{34}+\ln^{2}r_{14}+\ln^{2}r_{24}+\ln^{2}r_{34}]\frac{}{}\right\}\nonumber\\&\times(r_{12}r_{13}r_{14}r_{23}r_{24}r_{34})^{-2x_{E}(0)/3},
\end{eqnarray}

and

\begin{eqnarray}  &&\lim_{n\rightarrow0}
\langle:\phi_{a}\phi_{b}(r_{1})::\phi_{c}\phi_{d}(r_{2})::\phi_{e}\phi_{f}(r_{3})::\phi_{g}\phi_{h}(r_{4}):\rangle=A(0)f(\eta)
\nonumber
\\
&&\times\left\{D_{ab,cd,ef,gh}
+\frac{2}{3}u\left[\left(3\delta_{ab}D_{cd,ef,gh}\ln\frac{r_{23}r_{24}r_{34}}{(r_{12}r_{13}r_{14})^{2}}+{\rm
three\:\:terms}\right)\right.\right.\nonumber\\
&&-\left.\left.\left(\delta_{ab}\delta_{cd}D_{ef,gh}\ln\frac{r_{12}^{4}r_{13}r_{14}r_{23}r_{24}}{r_{34}^{2}}+{\rm
five\:\:terms}\right)\right]\right.\nonumber\\
&&+\left.\frac{8}{3}u^{2}\left\{\left[-\delta_{ab}\delta_{cd}D_{ef,gh}\left(\frac{}{}2(\ln
r_{13}+\ln r_{14})^{2}+2(\ln r_{23}+\ln r_{24})^{2}-
(2\ln r_{12}-\ln r_{34})^{2}\right.\right.\right.\right.\nonumber\\
&&+\left.\left.\left.\left.\ln(r_{13}r_{14}r_{23}r_{24})\ln\frac{r_{34}}{r_{12}^{2}}-5(\ln
r_{13}\ln r_{24}+\ln r_{14}\ln r_{24}+\ln r_{13}\ln r_{23}+\ln
r_{14}\ln
r_{23})\right)\right.\right.\right.\nonumber\\&&\left.\left.\left.+{\rm five\:\:terms}\frac{}{}\right]\right.\right.\nonumber\\
&&+\left.\left.\delta_{ab}\delta_{cd}\delta_{ef}\delta_{gh}\left[\frac{}{}-\left(\ln^{2}r_{12}+{\rm
five\:\:terms}\right)+\left(\ln r_{12}\ln r_{13}+\ln r_{12}\ln
r_{14}+\ln r_{13}\ln
r_{14}\right.\right.\right.\right.\nonumber\\
&&+\left.\left.\left.\left. {\rm three\:\:terms}\right)+(4\ln
r_{12}\ln
r_{34}+{\rm two\:\:terms})\frac{}{}\right]\right\}\right.\nonumber\\
&&+\left.\frac{16}{9}u^{3}\delta_{ab}\delta_{cd}\delta_{ef}\delta_{gh}\left\{\left(\frac{}{}3\ln^{2}r_{12}\ln(r_{13}r_{14}r_{23}r_{24})+4\ln^{3}r_{12}-6\ln^{2}r_{12}\ln
r_{34}\right.\right.
+{\rm five\:\:terms}\frac{}{}\right)\nonumber\\
&&-\left.\left.\left(\frac{}{}12\ln(r_{12}r_{13})\ln r_{24}\ln
r_{34}+{\rm five\:\:terms}
\right)\right.\right.+\left(\frac{}{}24\ln r_{12}\ln r_{13}\ln
r_{23}-30\ln r_{12}\ln r_{23}\ln
r_{24}\right.\nonumber\\
&&+\left.{\rm three\:\:terms}\frac{}{}\right)\\
&&+\left.\delta_{ab}\delta_{cd}\delta_{ef}\delta_{gh}\left[\frac{}{}-24A^{\prime\prime\prime}(0)+64B^{\prime\prime\prime}(0)-48C^{\prime\prime\prime}(0)
+8D^{\prime\prime\prime}(0)\frac{}{}\right]\right\}(r_{12}r_{13}r_{14}r_{23}r_{24}r_{34})^{-2x_{E}(0)/3}.\nonumber
\end{eqnarray}
Because of avoiding lengthy expression we didn't write explicit
form of all terms in the above equation, but it is easy to write
them by symmetry considerations.

\section{Summary}

 We have studied the correlation functions of self-avoiding walks
and derived their two-, three-, and four-point correlation
functions using the $O(n)$ model in the limit $n\to 0$. One can
directly check that the three- and four-point correlation
functions have the general properties of a logarithmic conformal
field theory, and that the logarithmic partner can be regarded as
the formal derivative of the ordinary fields (top field) with
respect to their conformal weight [9]. In this case, one can
consider the field $:\phi_a \phi_b:$ as the derivatives of field
$E$ with respect to $n$. We emphasize that the derivative with
respect to the scaling weight can be written in terms of the
derivative with respect to $n$. These properties enable us to
calculate any $N$-point correlation function that contains the
logarithmic field $:\phi_a\phi_b:$, in terms of the correlation
functions of the top fields. The general expression of the
correlation functions of the LCFT are given in Ref. [9]. Here, we
have determined the unknown constants in the logarithmic
correlation functions in terms of the details of the SAWs. It is
noted that the formal derivations with respect to the scaling
dimensions cannot predict the unknown constants in the quenched
averaged correlation functions of the local energy density
operators. The constants depend on the detail of the statistical
model.

Our analytical results can also be checked numerically. Our
analysis is valid in all dimensions below the upper critical
dimension. These results can be generalized to other problems,
such as
percolation, random phase sine-Gordon model, etc.\\

This paper is dedicated to Professor Ian Kogan.

\section{Appendix A}

Here, we present the details of the calculations for the operator
product expansion of $\Phi\cdot\Phi$, $E\cdot\Phi$ and
$\widetilde{E}_{ab}\cdot\Phi$. Using the definition $\Phi$, one
finds that:
\begin{eqnarray} \Phi.\Phi
 &=&(\sum_{ab}:\phi_a^2\phi_b^2:).(\sum_{cd}:\phi_c^2\phi_d^2:)\nonumber\\
 &=&4\sum_{abcd}(\delta_{ac}+\delta_{ac}\phi_{a})
\phi_{a}\phi_{c}\phi_{b}^2\phi_{d}^2
+4\sum_{abcd}(\delta_{ad}+\delta_{ad}\phi_{a})\phi_{a}\phi_{d}\phi_{b}^2\phi_{c}^2+4\sum_{abcd}(\delta_{bc}+\delta_{bc}\phi_{b})\phi_{b}\phi_{c}\phi_{a}^2\phi_{d}^2\nonumber\\
\quad&+&4\sum_{abcd}(\delta_{bd}+\delta_{bd}\phi_{b})\phi_{b}\phi_{d}\phi_{a}^2\phi_{c}^2+2\sum_{abcd}(\delta_{ac}+\delta_{ac}\phi_{a})(\delta_{ac}+\delta_{ac}\phi_{a})\phi_{b}^2\phi_{d}^2
\nonumber\\
\quad&+&2\sum_{abcd}(\delta_{ad}+\delta_{ad}\phi_{a})(\delta_{ad}+\delta_{ad}\phi_{a})\phi_{b}^2\phi_{c}^2+2\sum_{abcd}(\delta_{bc}+\delta_{bc}\phi_{b})(\delta_{bc}+\delta_{bc}\phi_{b})\phi_{a}^2\phi_{d}^2\nonumber\\
\quad
&+&2\sum_{abcd}(\delta_{bd}+\delta_{bd}\phi_{b})(\delta_{bd}+\delta_{bd}\phi_{b})\phi_{a}^2\phi_{c}^2+8\sum_{abcd}(\delta_{ac}+\delta_{ac}\phi_{a})(\delta_{ad}+\delta_{ad}\phi_{a})\phi_{b}^2\phi_{c}\phi_{d}\nonumber\\
\quad
&+&8\sum_{abcd}(\delta_{bc}+\delta_{bc}\phi_{b})(\delta_{bd}+\delta_{bd}\phi_{b})\phi_{a}^2\phi_{c}\phi_{d}+16\sum_{abcd}(\delta_{ac}+\delta_{ac}\phi_{a})(\delta_{bd}+\delta_{bd}\phi_{b})\phi_{a}\phi_{b}\phi_{c}\phi_{d}\nonumber\\
\quad
&+&16\sum_{abcd}(\delta_{ad}+\delta_{ad}\phi_{a})(\delta_{bc}+\delta_{bc}\phi_{b})\phi_{a}\phi_{b}\phi_{c}\phi_{d}+8\sum_{abcd}(\delta_{ac}+\delta_{ac}\phi_{a})(\delta_{bc}+\delta_{bc}\phi_{b})\phi_{a}\phi_{b}\phi_{d}^2\nonumber\\
\quad
&+&8\sum_{abcd}(\delta_{ad}+\delta_{ad}\phi_{a})(\delta_{bd}+\delta_{bd}\phi_{b})\phi_{a}\phi_{b}\phi_{c}^2+...,
\nonumber\\&=&24n^2+96nE+(8n+64)\Phi+... .\end{eqnarray}

  \begin{eqnarray}
E.\Phi&=&(\sum_{a}:\phi_{a}^{2}:).(\sum_{bc}:\phi_b^2\phi_c^2:)\nonumber\\
&=&4\sum_{abc}(\delta_{ab}+\delta_{ab}\phi_a)\phi_a\phi_b\phi_{c}^2
+4\sum_{abc}(\delta_{ac}+\delta_{ac}\phi_a)\phi_a\phi_c\phi_{b}^2\nonumber\\
\quad
&+&2\sum_{abc}(\delta_{ab}+\delta_{ab}\phi_a)(\delta_{ab}+\delta_{ab}\phi_a)\phi_{c}^2+2\sum_{abc}(\delta_{ac}+\delta_{ac}\phi_a)(\delta_{ac}+\delta_{ac}\phi_a)\phi_{b}^2\nonumber\\
\quad
&+&8\sum_{abc}(\delta_{ab}+\delta_{ab}\phi_a)(\delta_{ac}+\delta_{ac}\phi_a)\phi_b\phi_c,\nonumber\\
&=&8\Phi+(4n+8)E+... .\end{eqnarray} and,
\begin{eqnarray}
\widetilde{E}_{ab}.\Phi&=&(:\phi_{a}\phi_{b}:-\frac{\delta_{ab}}{n}\sum_{i}:\phi_{i}^{2}:).(\sum_{cd}:\phi_c^2\phi_d^2:)\nonumber\\
\quad
&+&2\sum_{cd}(\delta_{ac}+\delta_{ac}\phi_a)(\delta_{bc}+\delta_{bc}\phi_b)\phi_{d}^2+2\sum_{cd}(\delta_{ad}+\delta_{ad}\phi_a)(\delta_{bd}+\delta_{bd}\phi_b)\phi_{c}^2\nonumber\\
\quad
&+&4\sum_{cd}(\delta_{ac}+\delta_{ac}\phi_a)(\delta_{bd}+\delta_{bd}\phi_b)\phi_c\phi_d+4\sum_{cd}(\delta_{ad}+\delta_{ad}\phi_a)(\delta_{bc}+\delta_{bc}\phi_b)\phi_c\phi_d\nonumber\\
\quad
&-&\frac{\delta_{ab}}{n}(4n+8)\sum_{i}:\phi_{i}^2:+...,\nonumber \\
&=&-\frac{8}{n}\delta_{ab}\Phi+8\widetilde{E}_{ab}+...
.\end{eqnarray}

\section{Appendix B}
The detail of the derivation of the three- and four-points
correlation functions are presented in this Appendix. The
three-point correlation function of the field
$\widetilde{E}_{ab}(r)$ has the following explicit expression in
terms of the fields $:\phi_{a}\phi_{b}:$:
\begin{eqnarray}
&\hskip
-10mm\langle\widetilde{E}_{ab}(r_{1})\widetilde{E}_{cd}(r_{2})\widetilde{E}_{ef}(r_{3})\rangle=\left\langle(:\phi_{a}\phi_{b}(r_1):-\frac{\delta_{ab}}{n}\sum_{i}:\phi_{i}^{2}(r_1):)
(:\phi_{c}\phi_{d}(r_2):-\frac{\delta_{cd}}{n}\sum_{j}:\phi_{j}^{2}(r_2):)\right.\nonumber\\
&\hskip 50mm
\left.\times(:\phi_{e}\phi_{f}(r_3):-\frac{\delta_{ef}}{n}\sum_{k}:\phi_{k}^{2}(r_3):)\right\rangle
.\end{eqnarray} Tensorial structure of the right-hand side of the
above equation has the following terms:
\begin{eqnarray}
\langle:\phi_{a}\phi_{b}::\phi_{c}\phi_{d}::\phi_{e}\phi_{f}:\rangle&\sim&
D_{ab,cd,ef},\nonumber\\
\sum_{i}\langle:\phi_{a}\phi_{b}::\phi_{c}\phi_{d}::
\phi_{i}^{2}:\rangle&\sim&4D_{ab,cd},\nonumber\\
\sum_{ij}\langle:\phi_{a}\phi_{b}::\phi_{i}^{2}::\phi_{j}^{2}:\rangle&\sim&8\delta_{ab},\nonumber\\
\sum_{ijk}\langle:\phi_{i}^{2}::\phi_{j}^{2}::\phi_{k}^{2}:\rangle&\sim&8n,
\end{eqnarray}
where
$D_{ab,cd,ef}=\delta_{ac}D_{bd,ef}+\delta_{ad}D_{bc,ef}+\delta_{ae}D_{cd,bf}+\delta_{af}D_{be,cd}
$. According to Eq.(28) and the above equations:
\begin{eqnarray}\label{3point}
 \langle
\widetilde{E}_{ab}(r_{1})\widetilde{E}_{cd}(r_{2})\widetilde{E}_{ef}(r_{3})\rangle&=&
\left[D_{ab,cd,ef}-\frac{4}{n}(\delta_{ab}D_{cd,ef}+\delta_{cd}D_{ab,ef}+\delta_{ef}D_{ab,cd})+\frac{16}{n^2}\delta_{ab}
\delta_{cd}\delta_{ef}\right]\nonumber\\
\quad&\times& A(n)(r_{12}r_{13}r_{23})^{-x_{\widetilde{E}}(n)}.
\end{eqnarray}
Other three-point correlation functions which are used in deriving
Eqs.(\ref{sadegh},31), obtain by the same method:
\begin{eqnarray}
\langle \widetilde{E}_{ab}(r_{1})E(r_{2})E(r_{3})\rangle&=&0,\nonumber\\
\langle
E(r_{1})E(r_{2})E(r_{3})\rangle&=&8nC(n)(r_{12}r_{13}r_{23})^{-x_{E}(n)},
\nonumber\\\langle
\widetilde{E}_{ab}(r_{1})\widetilde{E}_{cd}(r_{2})E(r_{3})\rangle&=&\left[4D_{ab,cd}-\frac{8}{n}\delta_{ab}
\delta_{cd}\right]B(n)(r_{13}r_{23})^{-x_{E}(n)}r_{12}^{x_{E}(n)-2x_{\widetilde{E}}(n)}.
\end{eqnarray}

 Four-point correlation function of four scaling fields $\varphi_1$, $\varphi_2$, $\varphi_3$
 and $\varphi_4$ with scaling dimensions $\Delta_1$, $\Delta_2$, $\Delta_3$ and
 $\Delta_4$, respectively, has the form:
 \begin{eqnarray}
\langle\varphi_1(r_1)\varphi_2(r_2)\varphi_3(r_3)\varphi_4(r_4)\rangle=\prod_{1\leq
i<j\leq4} r_{ij}^{\Delta_{ij}}f(\eta),
 \end{eqnarray}
with
$\Delta_{ij}=\frac{1}{3}\sum_{k=1}^4\Delta_k-\Delta_i-\Delta_j$
and $f(\eta)$ is the unknown function of cross ratio
$\eta=\frac{r_{_{12}}r_{_{34}}}{r_{_{13}}r_{_{24}}}$.

  According to above equation some key functions that are used for deriving the four-point
correlation functions are as follows:
\begin{eqnarray}
&&\langle
\widetilde{E}_{ab}(r_{1})E(r_{2})E(r_{3})E(r_{4})\rangle=0,\nonumber\\
&&\langle
E(r_{1})E(r_{2})E(r_{3})E(r_{4})\rangle=A(n)f_{1}(\eta)(r_{12}r_{13}r_{14}r_{23}r_{24}r_{34})^{-2x_{E}(n)/3}(48n+12n^2),\nonumber\\
&&\langle
\widetilde{E}_{ab}(r_{1})\widetilde{E}_{cd}(r_{2})\widetilde{E}_{ef}(r_{3})E(r_{4})\rangle=B(n)f_{2}(\eta)(r_{12}r_{13}r_{23})^{-x_{\widetilde{E}}(n)+x_{E}(n)/3}
(r_{14}r_{24}r_{34})^{-2x_{E}(n)/3}\nonumber\\
&\times&\left[6D_{ab,cd,ef}-\frac{24}{n}(\delta_{ab}D_{cd,ef}+\delta_{cd}D_{ab,ef}+\delta_{ef}D_{ab,cd})+\frac{96}{n^2}\delta_{ab}\delta_{cd}\delta_{ef}\right],
\nonumber\\
&&\langle
\widetilde{E}_{ab}(r_{1})\widetilde{E}_{cd}(r_{2})E(r_{3})E(r_{4})\rangle=C(n)f_{3}(\eta)\left[(2n+24)D_{ab,cd}-\left(4+\frac{48}{n}\right)\delta_{ab}\delta_{cd}\right]\nonumber\\
&\times& (r_{12}^{-4x_{\widetilde{E}}(n)/3+2x_{E}(n)/3})
(r_{34}^{2x_{\widetilde{E}}(n)/3-4x_{E}(n)/3})(r_{13}r_{14}r_{23}r_{24})^{-x_{\widetilde{E}}(n)/3-x_{E}(n)/3},\nonumber\\
&&\langle
\widetilde{E}_{ab}(r_{1})\widetilde{E}_{cd}(r_{2})\widetilde{E}_{ef}(r_{3})\widetilde{E}_{gh}(r_{4})\rangle=D(n)f_{4}(\eta)
(r_{12}r_{13}r_{14}r_{23}r_{24}r_{34})^{-2x_{\widetilde{E}}(n)/3}\nonumber\\
\quad&\times&\left[D_{ab,cd,ef,gh}-\frac{6}{n}\alpha_{abcdefgh}+\left(\frac{24}{n^2}-\frac{2}{n}\right)\beta_{abcdefgh}+\left(\frac{12}{n^2}-\frac{144}{n^3}
\right)\delta_{ab}\delta_{cd}\delta_{ef}\delta_{gh}\right].
\end{eqnarray}
where
\begin{eqnarray}
D_{ab,cd,ef,gh}&=&\delta_{ac}(\delta_{bd}D_{ef,gh}+D_{bd,ef,gh})+
\delta_{ad}(\delta_{bc}D_{ef,gh}+D_{bc,ef,gh})\nonumber\\
&+& \delta_{ae}(\delta_{bf}D_{cd,gh}+D_{bf,cd,gh})
+\delta_{af}(\delta_{be}D_{cd,gh}+D_{be,cd,gh})\nonumber\\&+&
\delta_{ag}(\delta_{bh}D_{cd,ef}+D_{bh,cd,ef})+
\delta_{ah}(\delta_{bg}D_{cd,ef}+D_{bg,cd,ef}),\nonumber\\
\alpha_{abcdefgh}&=&\delta_{ab}D_{cd,ef,gh}+\delta_{cd}D_{ab,ef,gh}+\delta_{ef}D_{ab,cd,gh}+\delta_{gh}D_{ab,cd,ef},\nonumber\\
\beta_{abcdefgh}&=&\delta_{ab}\delta_{cd}D_{ef,gh}+\delta_{ab}\delta_{ef}D_{cd,gh}+\delta_{ab}\delta_{gh}D_{cd,ef}\nonumber\\
&+&\delta_{cd}\delta_{ef}D_{ab,gh}+\delta_{cd}\delta_{gh}D_{ab,ef}+\delta_{ef}\delta_{gh}D_{ab,cd}.
\end{eqnarray}


\end{document}